# Energetics, electronic structure and electric polarization of basal stacking faults in wurtzite GaN and ZnO

Abdesamed Benbedra[a], Said Meskine[a], Abdelkader Boukortt[a], Roland Hayn[b], Michael Texier[b], Olivier Thomas[b], and Thomas W. Cornelius[b]

[a] Laboratoire d'Elaboration et Caractérisation Physico Mécanique et Métallurgique des Matériaux (ECP3M), Université Abdelhamid Ibn Badis, Route Nationale N°11, Kharouba, 27000 Mostaganem, Algérie
[b] Aix-Marseille Université, Univ de Toulon, CNRS, IM2NP-UMR 7334, Marseille, France

**Abstract**

We investigate the effect of basal-plane stacking faults on the structural, electronic, and polarization properties of wurtzite GaN and ZnO. This theoretical study is performed within density-functional theory (DFT) using periodic hexagonal supercells. Both formation energies and band structures are obtained by means of total-energy calculations. The type-I stacking fault is observed to have the lowest formation energy, followed by type-II and finally the extrinsic stacking fault. In order to overcome the inherent shortcoming of DFT in reproducing band gaps, the generalized-gradient approximation is used in combination with the modified Becke-Johnson functional. It is shown that all stacking faults studied maintain a direct gap whose value is lower than that in the ideal defect-free crystals. The lowering in the band gap allows the creation of quantum-well regions at wurtzite/zincblende interfaces. In addition, we provide a consistent set of polarization parameters derived from the Berry-phase method. We find a trend of decreasing (increasing) spontaneous polarization and piezoelectric coefficient (polarization charge) in going from type-I to type-II to extrinsic stacking faults. We compare our results to experimental and theoretical data available from the literature and explain the observed trends in terms of the properties of the wurtzite and zincblende polytypes of both materials.

## I. Introduction

At present, there is a regain of interest in wide band gap semiconductors. Those with direct band gap like GaN or ZnO, are not only very interesting for optoelectronics, but also in power electronics [1-3]. Nanostructuration gives new functionalities and opens a huge field of fundamental and applied research [4-7]. For instance, nanowires are thought to be an essential ingredient of future nanoelectronics. Especially compelling in that respect are polar crystals with wurtzite structure since controlling the inherent electric field gives rise to an additional leverage for manipulating electric currents. Both wurtzite GaN and ZnO are usually grown as thin films on mismatched substrates such as sapphire and SiC [8-12]. Such systems are characterized by residual strains at the interface, invariably chosen to be the (0001) plane (or *c*-plane) [13]. The origin of these strains is the difference in the in-plane lattice constant of the film and substrate layers [14]. The mismatch-induced strains can be very high, thus resulting in the formation of a large density of extended structural defects. These include, but not limited to, dislocations [15,16], inversion domain boundaries [17,18] and stacking faults [19-22]. These defects can also be generated during growth on crystallographic planes without strain relaxation [23].

Stacking faults constitute a common type of extended two dimensional defects in GaN and ZnO, which are expected to have an impact on physical properties and hence on device performance. Typically, they can be generated either through the dissociation of a full dislocation into two partial dislocations, whose separation is related to the energy of the stacking fault, or the nucleation and propagation of isolated partial dislocations, or the condensation of isolated point defects (i.e. interstitial atoms or vacancies) leading to the insertion or the removing of an atomic bilayer within the crystalline stacking sequence [24]. High-resolution transmission electron microscopy (TEM) revealed the existence of such defects even in GaN (ZnO) films epitaxially grown on GaN (ZnO) substrates [25,26]. There are three types of stacking faults in hexagonal crystals, depending on the plane on which the defect occurs. These are prism-, pyramidal- and basal-plane stacking faults [27]. The latter are the subject of the present paper. There are several sub-types of basal-plane stacking faults which will all be studied. They can be understood by introducing different numbers of zincblende or cubic layers into the wurtzite stacking sequence.

Most theoretical investigations so far have focused on the energetics and electronic structure of basal stacking faults in wurtzite III-nitrides and II-oxides. Surprisingly, the influence of stacking faults on the polarization and the band offsets has not yet been sufficiently studied in comparison to their technological importance. Indeed, wurtzite materials are found to be highly polarized [28-31], and a study of the effect of stacking faults on polarization-related quantities is of central importance. This lack motivated the present work, in which we study the structural, electronic and polarization properties of basal-plane stacking faults in wurtzite GaN and ZnO via first-principles methods. This amounts to calculating structural parameters, formation energies, band gaps, band offsets, spontaneous polarizations, polarization charges and piezoelectric coefficients. The main goal is to investigate to what extent these quantities can be affected by the presence of different types of basal stacking faults. We compare the simulation results with available experimental data showing very good agreement with some of them which helps to assess their quality. Furthermore, we detect a simple rule of thumb which connects the formation energies, band gaps, and polarizations of the different stacking faults simply with the concentration of cubic layers.

The paper is structured as follows. Section II describes the theoretical approach: in section II.1 we give an overview of the basal-plane stacking faults in hexagonal crystals. Section II.2 deals with the supercell representation of all types of stacking faults studied here. Section II.3 introduces the numerical details of the simulation. Section III contains the main results of the paper: sections III.1-III.3 present the computational results and discussion of the structural optimization, electronic structure and electric polarization, respectively. Finally, in section IV we summarize the results and present our conclusions in section V.

## II. Theoretical approach

## II.1 Basal stacking faults in hexagonal crystals

The wurtzite structure in which GaN and ZnO crystallize is formed by the stacking of atoms in the sequence ..AaBbAaBb.. along the [0001] direction [32]. The zincblende structure, on the other hand, can be described by the stacking sequence ..AaBbCcAaBbCc.. along the [111] direction [32]. Here, upper-case and lower-case letters correspond to cations (Ga and Zn) and anions (N and O), respectively. The letters Aa, Bb and Cc indicate three different positions of basal or *c*-plane atomic bilayers. The *x*- and *y*- coordinates of atoms at the Aa, Bb and Cc layers are respectively (1/3,2/3), (2/3,1/3) and (0,0). In the wurtzite structure, every bilayer is surrounded by two identical bilayers (e.g. Aa<u>Bb</u>Aa). In this case Aa-Bb bonds are called hexagonal bonds. In the zincblende structure, the two bilayers surrounding a given bilayer are different (e.g. Aa<u>Bb</u>Cc), and the Aa-Bb bonds are called cubic bonds [33]. Stacking faults can then be viewed as local deviations of the perfect wurtzite stacking sequence. In other words, basal stacking faults are embedded cubic units surrounded by a hexagonal matrix [34].

There are three types of basal-plane stacking faults reported in the literature [23,33,35,36]: two intrinsic stacking faults of type-I and type-II, and one extrinsic stacking fault. The type-I stacking fault originates from one violation of the wurtzite stacking rule. Specifically, it leads to a stacking sequence of ....AaBbCcBbCc.... or ....AaBbAaCcAaCc..... This stacking fault introduces one cubic layer with cubic bonds above and below. In our study we have to impose periodic boundary conditions which can only be done by introducing two stacking faults. In the case of type-I stacking fault that leads to the stacking sequence of ...AaBbCcBbAa... (see Sec. II.2). The type-II stacking fault results from two violations of the normal stacking rule leading to the sequence ....AaBbCcAaCc... (or ...AaBbAaCcBbCc... ). This stacking fault introduces two cubic layers. Finally, the extrinsic stacking fault contains an additional Cc layer inserted in the midst of the original stacking sequence: ...AaBbAaBbCcAaBbAaBb.... It introduces three cubic layers at the faulted interface.

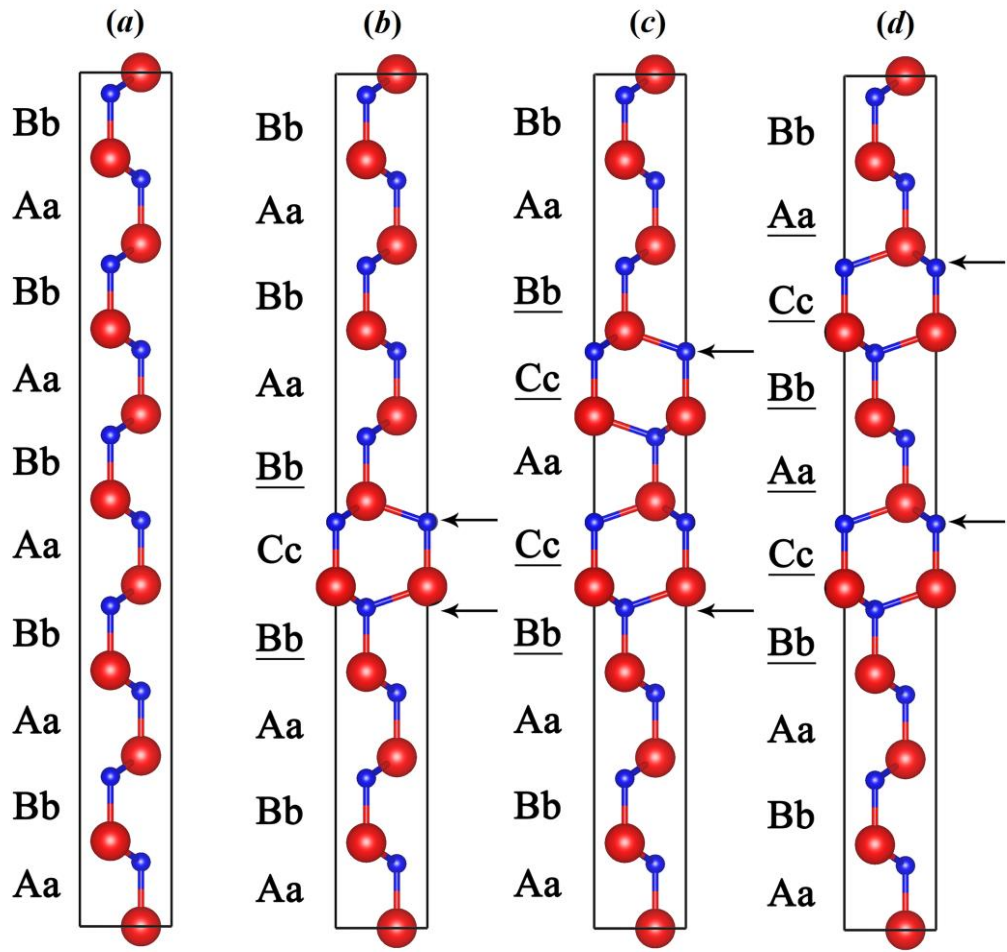

**Figure 1** Supercell models representing the studied structures for (*a*) ideal crystal, (*b*) type-I, (*c*) type-II and (*d*) extrinsic stacking faults. The position of the defects is indicated by an arrow, and the cubic layers are indicated by underlined letters. Red and blue spheres correspond to Ga/Zn and N/O atoms, respectively. We make use of the VESTA software [37] to visualize these structures.

## II.2 Supercells choice

In order to model the stacking faults, we use periodic supercells containing 20 atoms, i.e. 10 double-layers, in a construction of 1×1×5 wurtzite unit cells. The different types of stacking faults are achieved by a proper manipulation of specific atomic positions applied to the original wurtzite stacking sequence. As an example, the *x*- and *y*-coordinates of atoms in a Bb layer are changed from (2/3, 1/3) to (0,0) to create a Cc layer. Note that we did not highlight the *z*-coordinates because they are irrelevant, since they are kept fixed during the creation of the new Cc layer. In this way we obtain a type-I stacking fault. The adopted supercells of the stacking faults in both GaN and ZnO are presented in Figure 1. Due to imposing periodic boundary conditions, the beginning and the end of the supercells along the *c*-axis have to match. This has the effect of doubling both the number of cubic bonds and stacking faults. Indeed, the type-I stacking-fault supercell shown in Fig. 1(*b*) contains two cubic bonds instead of one. The same remark can be deduced for type-II and extrinsic stacking-fault supercells, for which there are four and six cubic bonds respectively (instead of two and three). This effect must be taken into account when calculating the formation energy of stacking faults [see Eq. (1) below]. Similarly, each supercell has two stacking faults, as indicated by the arrows in Fig. 1.

## II.3 Calculation details

The physical properties of basal-plane stacking faults (henceforth called stacking faults for brevity) in wurtzite GaN and ZnO are calculated within the framework of density-functional theory [38,39]. We use the WIEN2k package [40] with the generalized-gradient approximation (GGA) [41] to treat the exchange-correlation interactions. The modified Becke-Johnson (mBJ) approach [42,43] is chosen for dealing with the electronic properties, since it was successfully applied for studying such properties in many materials. The Kohn-Sham equations [44] are solved self-consistently employing the full-potential linearized-augmented plane-wave (FP-LAPW) algorithm [45,46]. The wavefunctions cutoff-parameters $l_{max}$ and $R_{MT}^{min} k_{max}$ are set to 10 and 7, respectively, where $l_{max}$, $R_{MT}^{min}$ and $k_{max}$ are defined in Ref. [40]. The first Brillouin zone is sampled non-uniformly using a 4×4×8 mesh. The energy that separates core and valence states is chosen such that the following orbitals are treated as valence electrons: Ga (3d 4s 4p), N (2s 2p), Zn (3d 4s) and O (2s 2p).

The structural equilibrium parameters and electronic properties for each stacking fault are studied by means of standard total-energy calculations. Because energy differences associated with stacking faults are quite small (in meV), a special care should be taken to obtain accurate values of energy. This is why in our calculation the energy convergence criterion is reduced down to $10^{-6}$ Ry, which is less than the default value by a factor of 100. To determine the internal parameter of the wurtzite structure (i.e. the *z*-coordinate of the anion), we carry out a relaxation process by minimizing the Hellman-Feynman forces [47,48] acting on atomic sites, such that the convergence condition on the forces is $10^{-6}$ Ry/Bohr. Finally, the code BerryPI [49] is used to evaluate polarization properties as finite differences following the guidance of the Berry-phase theory [50,51].

# III. Results and discussion

## III.1 Energetics and structural optimization

### a) Structural parameters

We begin our study by determining the equilibrium structure of each stacking fault. In doing so, we calculate the energy of the system on a grid of values of supercell volume and lattice-constant ratio $c/a$ [52]. Also, the atomic positions along the hexagonal axis ($c$-axis) are allowed to relax. The equilibrium lattice parameters are those who minimize the total energy [53]. The results of this calculation, namely the lattice constants $a$ and $c$ as well as the internal parameter $u$, are gathered in Table 1 for the ideal defect-free structures and stacking faults structures of wurtzite GaN and ZnO. Our calculated lattice parameters show good agreement with the ones determined from X-ray-diffraction measurements [54,55]. It can be seen that the structural relaxation has a negligible effect on the lattice parameters, as their values are practically the same for all the structures studied. In particular, stacking faults lying at the basal plane do not seem to perturb the value of the internal parameter. This is in accordance with the previous studies of stacking faults in GaN and ZnO [33,36].

Another structural parameter characterizing stacking faults, also presented in Table 1, is their thickness $\ell$. Intuitively, this is defined as the length of one cubic bond (which is $0.5c$ [56]) multiplied by the number of these bonds contained in the stacking fault. Hence, the thickness of type-I, -II, and extrinsic basal stacking faults would be $0.5c$, $c$ and $1.5c$, respectively [56]. Compared to the work of Lähnemann, et *al.* (see Table 1), who obtained the values of $\ell$ from the profile of the microscopic electrostatic potential, our results are lower by less than 2 Å.

**Table 1** Calculated equilibrium lattice constants $a$ and $c$, internal parameter $u$ (in units of $c$), and stacking-fault thickness $\ell$ for the ideal and stacking faults structures of wurtzite GaN and ZnO. Experimental [54,55] and theoretical [25] values from the literature (when available) are in parenthesizes.

|  | Ideal | Type-I | Type-II | Extrinsic |
|---|---|---|---|---|
| **GaN** | | | | |
| $a$ (Å) | 3.23 (3.20[a]) | 3.23 | 3.23 | 3.23 |
| $c$ (Å) | 5.25 (5.22[a]) | 5.25 | 5.25 | 5.25 |
| $u$ | 0.377 (0.376[a]) | 0.377 | 0.377 | 0.377 |
| $\ell$ (Å) | 0 | 2.63 (4.41[b]) | 5.25 (7.10[b]) | 7.88 (9.6[b]) |
| **ZnO** | | | | |
| $a$ (Å) | 3.29 (3.250[c]) | 3.29 | 3.29 | 3.29 |
| $c$ (Å) | 5.29 (5.204[c]) | 5.30 | 5.30 | 5.30 |
| $u$ | 0.381 (0.383[c]) | 0.380 | 0.380 | 0.380 |
| $\ell$ (Å) | 0 | 2.65 | 5.29 | 7.94 |

[a] Ref. [54], [b] Ref. [25], [c] Ref. [55]

## b) Formation energy

Stacking faults are not the ground-state of the material, they cost energy, which we refer to as formation energy $E_f$. The latter is given by the following formula [33,36]:

$$E_f = E^{SF} - E^{Ref}, \qquad (1)$$

with $E^{SF}$ is the total energy of the supercell containing the stacking fault and $E^{Ref}$ is that of the reference defectless supercell. As already mentioned, due to periodic boundary conditions the number of cubic bonds for each stacking fault is doubled. To account for this one should divide the result of Eq. (1) by two. Table 2 gives the calculated stacking-fault formation energies. Using smaller (12 atoms) and larger (32 atoms) supercells produces the same results as the 20-atom supercell. All stacking faults are found to have a very low formation energy, implying that these defects may form very easily in GaN and ZnO grown on mismatched substrates.

We compare our theoretical results with previous simulations. For stacking faults in GaN, our calculations show reasonable agreement with the results of Stampfl and Van de Walle [36], obtained from DFT using 16- and 18-atom supercells. The type-I and type-II stacking-fault energies from Batyrev *et al*. [57] agree well with our results. As for ZnO, there is an excellent agreement between the DFT results of Yan *et al*. [33] and the present work for all types of stacking faults. Nakamura *et al*. [58] computed the energy of type-II stacking fault associated with the dissociation of 60° dislocations into two Shockley partial dislocations, they found a formation energy 10 meV higher than our value for the same defect. Regarding experimentally obtained values, the energy of type-I stacking-fault of GaN reported by Takeuchi and Suzuki [59] (deduced from dissociation width measurements by TEM) compares extremely well with our result. Nakamura *et al*. [58] also estimated the formation energy of ZnO using a model based on the measured separation of two partial dislocations, which resulted in a value of 0.14 J/m² or 82.6 meV/uca. The authors did, however, not specify the type of the studied fault. Assuming it is the extrinsic type this value is significantly larger than our calculated energy of 47.6 meV/uca. This discrepancy may be attributed to uncertainties related to the measurement of the separation of the two partial dislocations or to the validity of the theoretical model used to estimate $E_f$.

**Table 2** Stacking-fault formation energies of wurtzite GaN and ZnO. The energy is given in meV per unit cell area of the faulted basal plane (uca) and in mJ/m². The conversion factor between the two units is 0.56 for GaN and 0.59 for ZnO. In brackets are previous literature data taken from: Stampfl and Van de Walle [36] (simulation), Batyrev *et al*. [57] (simulation), Takeuchi and Suzuki [59] (experiment), Yan *et al*. [33] (simulation), and Nakamura *et al*. [58] (simulation and experiment). See text for discussion.

|  | Type-I | Type-II | Extrinsic |
| --- | --- | --- | --- |
| **GaN** | | | |
| $E_f$ (meV/uca) | 10.5 (10[a], 12.32[b], 11.2[c]) | 21.5 (24[a], 25.20[b]) | 32.9 (38[a]) |
| $E_f$ (mJ/m²) | 18.8 (17.9[a], 22[b], 20[c]) | 38.4 (42.9[a], 45[b]) | 58.8 (67.9[a]) |
| **ZnO** | | | |
| $E_f$ (meV/uca) | 16 (15[d]) | 31.9 (31[d], 43.07[e]) | 47.6 (47[d], 82.6[e]) |
| $E_f$ (mJ/m²) | 27.1 (25.4[d]) | 54.01 (52.5[d], 73[e]) | 80.7 (79.7[d], 140[e]) |

[a] Ref. [36], [b] Ref. [57], [c] Ref. [59], [d] Ref. [33], [e] Ref. [58]

As can be readily seen from Table 2, the formation energy for both compounds is the lowest for type-I, followed by type-II and then extrinsic stacking fault. Specifically, stacking fault of type-II is twice as high in energy compared to type-I, while extrinsic stacking fault is three times larger. This trend can be explained by the fact that the formation energy is proportional to the number of cubic bonds at the stacking fault [33]. As such, the energy needed to form one cubic bond is equal to that of type-I stacking fault (since this fault contains only one cubic bond). The same reasoning applies to the other types as well: the energy of the type-II (extrinsic) stacking fault is two (three) times the energy of one cubic bond.

In order to shed more light on $E_f$, we calculate the energy difference between the zincblende and wurtzite polytypes. This difference is calculated to be 11.87 meV for GaN and 19.01 meV for ZnO. We note that both values are very close to the formation energy of type-I stacking fault. This is to be expected since the wurtzite-zincblende transformation involves the changing of the stacking sequence from AaBb to AaBbCc (or equivalently the creation of one cubic bond), so the energy difference of the two competing phases is consistent with the trend observed for the stacking-fault formation energies [59].

## III.2 Electronic structure

### a) Band gap and band offsets

Next, we investigate the electronic structure of basal stacking faults. The GGA and mBJ approximations are both employed to model the exchange-correlation part of the total-energy functional. The band structures calculated at the points of high symmetry are illustrated in Figure 2. For a reason described below, we only show the mBJ results. A detailed analysis of the band structure reveals that stacking faults introduce a downward (upward) shift at the conduction-band-minimum $E_C$ (valence-band-maximum $E_V$), resulting in a narrowing of the band gap $E_g$, In addition, we find that all stacking faults show a direct gap at point Γ. These findings are consistent with the previous works concerning stacking-faults electronic structure in GaN and ZnO [33,36]. The computed band gaps at the level of GGA and mBJ are listed in Table 3 for all GaN and ZnO stacking faults as well as the perfect structures. Also included are the mBJ values of the conduction and valence band offsets (discontinuities) $\Delta E_C$ and $\Delta E_V$. The conduction band offset is calculated by the difference:

$$\Delta E_C = E_C^{Ref} - E_C^{SF}, \qquad (2)$$

where $E_C^{Ref}$ and $E_C^{SF}$ are the conduction-band-minimum of the perfect and defected crystals, respectively. A similar expression exists for $\Delta E_V$, for which the valence-band-maxima $E_V^{Ref}$ and $E_V^{SF}$ are used. As will be shown later on, these quantities are of interest for the design of GaN- and ZnO-based optoelectronic devices, since band offsets affect the confinement energies of electrons and holes [60].

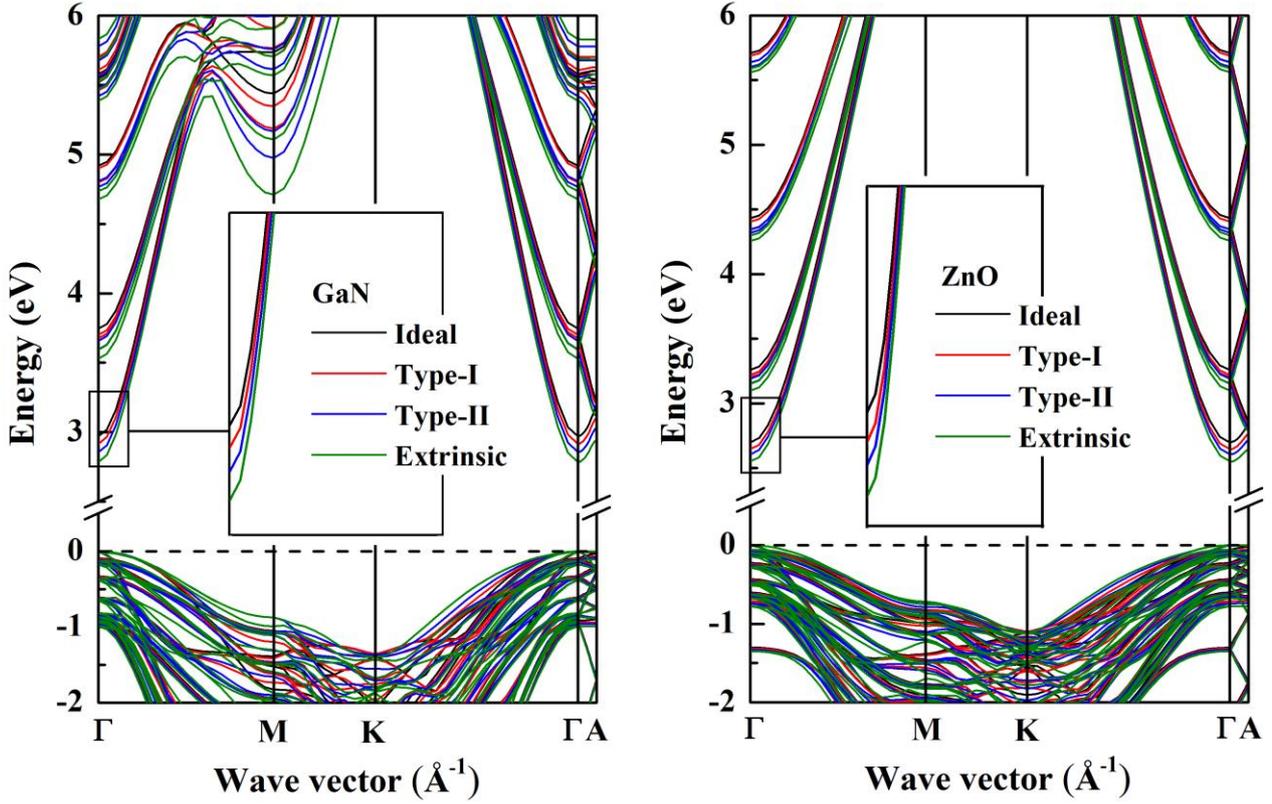

**Figure 2** Band structures for the ideal and stacking faults structures of wurtzite GaN and ZnO obtained using the mBJ approximation. The vertical energy scale is kept fixed for both materials. The origin of energies is the top of valence band (the dashed horizontal line). The values of the band gaps and band offsets are extracted from these curves. The insets give a closer look at the conduction band for all the structures studied.

**Table 3** Band gap $E_g$ as well as band offsets $\Delta E_C$ and $\Delta E_V$ within the mBJ approximation for the ideal and stacking faults structures of wurtzite GaN and ZnO. The GGA values for the gap are in parenthesizes.

|  | Ideal | Type-I | Type-II | Extrinsic |
|---|---|---|---|---|
| **GaN** | | | | |
| $E_g$ (eV) | 2.97 | 2.92 | 2.86 | 2.78 |
|  | (1.68) | (1.63) | (1.56) | (1.54) |
| $\Delta E_C$ (eV) | 0 | 0.045 | 0.096 | 0.16 |
| $\Delta E_V$ (eV) | 0 | -0.011 | -0.019 | -0.028 |
| **ZnO** | | | | |
| $E_g$ (eV) | 2.67 | 2.62 | 2.58 | 2.53 |
|  | (0.75) | (0.72) | (0.69) | (0.67) |
| $\Delta E_C$ (eV) | 0 | 0.048 | 0.070 | 0.094 |
| $\Delta E_V$ (eV) | 0 | -0.002 | -0.021 | -0.046 |

Our results show that the gap is generally the largest for type-I, the smallest for extrinsic, and intermediate for type-II stacking fault. Using the local-density approximation, Stampfl and Van de Walle [36] have found that the bandgap reduces by 0.02 eV for GaN type-I stacking fault. From our data in Table 3, we obtain a decrease of 0.05 eV. To the best of our knowledge, there are no published studies for the energy gap to compare our results with, for which this work seems to be the first one to report values. Both $\Delta E_C$ and $\Delta E_V$ are observed to be smaller for type-I stacking fault, but larger for the extrinsic one. We also note that the conduction and valence band offsets are of opposite sign, with the dominant offset being the conduction part. Most of the available data for the band offsets is for wurtzite/zincblende superlattice interfaces, rather than the stacking faults themselves. Yan *et al*. [33] have shown that $\Delta E_C$ for stacking faults in ZnO ranges from 0.02 to 0.04 eV, but have not reported explicit values. Compared to these calculations, our values are in rather good agreement.

A by-product of the calculation of the zincblende/wurtzite energy difference reported above is the values of the band gap for each phase. In the case of zincblende, we obtain the following values: 2.74 and 2.51 eV for GaN and ZnO, respectively. The corresponding gaps for the wurtzite system are listed in the second column of Table 3. In going from type-I to type-II to extrinsic stacking faults, the crystalline structure gets closer and closer to the zincblende phase. Therefore, the band gaps associated with the stacking faults lie between the limiting values of the wurtzite and zincblende polytypes. From Fig. 1 we can read out a concentration of cubic layers of 20, 40, and 60 % for type-I, type-II, and extrinsic stacking faults in our chosen supercells, respectively. One remarks from Table 3 that the band-gap change is in a rather good approximation simply proportional to the concentration of cubic bonds, both for GaN and ZnO. It would be interesting to test this hypothesis also for larger supercells with a smaller concentration of cubic bonds.

The experimental band gap of wurtzite GaN (ZnO) is around 3.5 eV [61] (3.4 eV [62]). It is evident that the gap values for the perfect crystals obtained by GGA are severely underestimated compared to those found experimentally. In particular, for the case of ZnO the discrepancy between theory and experiment is about 76%. It is well established that the GGA functional fails to reproduce the band structure of semiconductors by underestimating their band gap [63], often by several electronvolts. To surmount this difficulty, we use the mBJ functional [64], which provides much more accurate and reliable results for the energy gap. This fact is demonstrated by our results (see Table 3), where mBJ-corrected gaps are found to be in better agreement with experiment than GGA.

**b) Stacking faults as quantum wells**

Stacking faults considered here can be pictured as thin zincblende layers surrounded by a wurtzite lattice [65], so that one speaks of zincblende/wurtzite interfaces. Because of the lowering in the band gap discussed earlier, these stacking faults give rise to quantum-well-like regions as already proposed by many authors [34,66,67]. The band profile of such quantum wells is shown in Figure 3 for GaN and ZnO. Electrons in the conduction band with small energy gap (i.e. in the zincblende layer) will be trapped in this region. The situation is similar for holes in the valence band. In GaN and ZnO semiconductor devices, these quantum wells act as a trap center for charge carriers and may lead to photoluminescence emission lines [56,68-70]. We point out that the confinement energy-barrier, or the height of the quantum well, is fully determined by the band offsets. As a consequence, the quantum well associated with the extrinsic stacking fault has the highest barrier, as shown in Fig. 3.

Let us point to the importance of stacking faults as traps for charge carriers. We consider ZnO as an example. It is well known that most grown ZnO crystals are n-doped due to the presence of oxygen vacancies which is hard to avoid in real systems. As it is visible in Fig. 3 each stacking fault in ZnO gives rise to a potential well. On the other hand, it is well known from basic quantum mechanics that each quantum well of even small depth has at least one bound state which will capture the doped electrons.

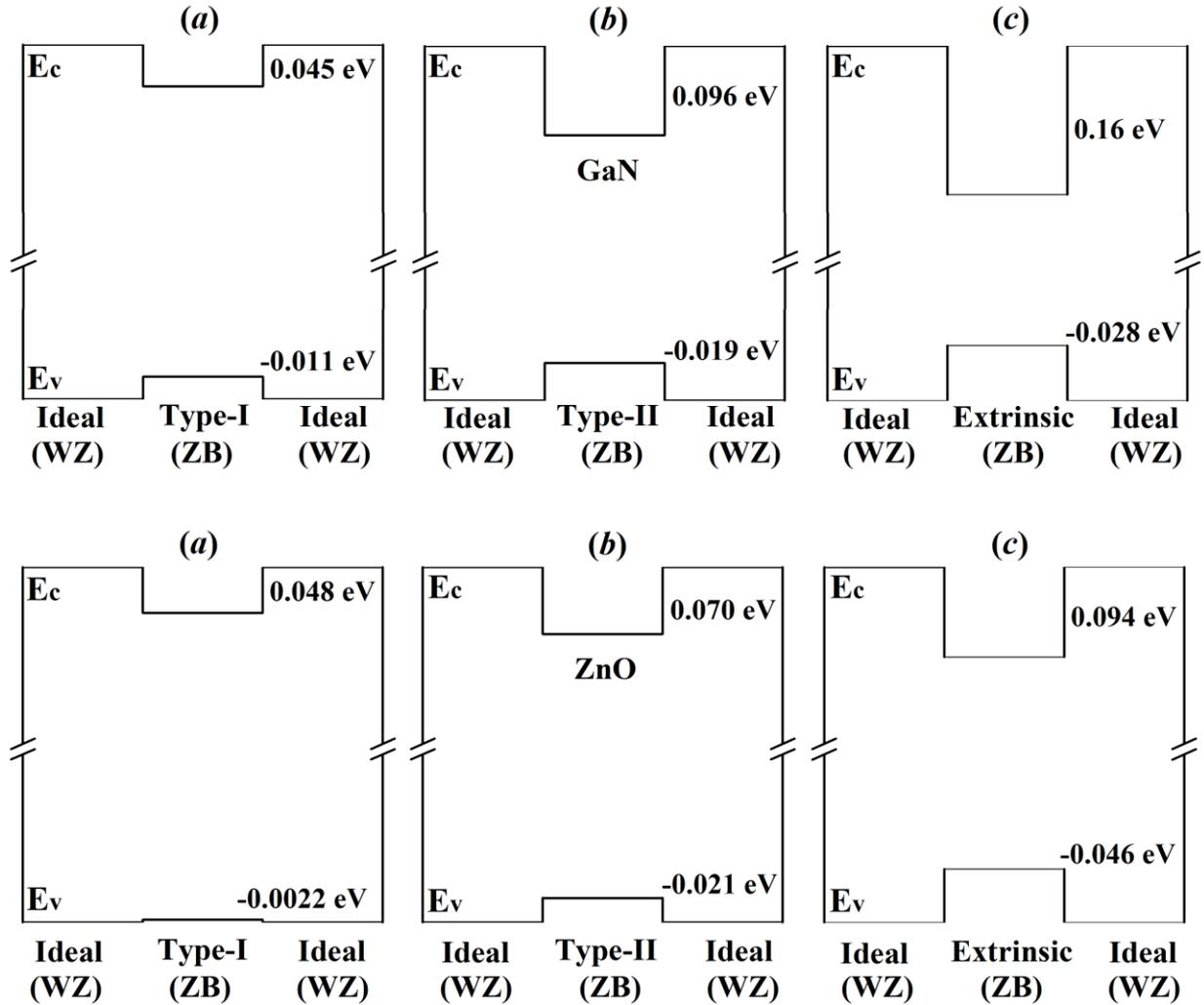

**Figure 3** Schematic plot of the band profile of quantum wells associated with (*a*) type-I, (*b*) type-II and (*c*) extrinsic stacking faults in GaN (up) and ZnO (down). "Ideal" refers to the wurtzite (WZ) side of the interface while the stacking faults refer to the zincblende (ZB) side. The values of the conduction and valence band-offsets are also shown.

### III.3 Electric polarization

**a) Spontaneous polarization and polarization charge**

The polarization of any wurtzite crystal is the sum of spontaneous and piezoelectric components. The first component is due to the intrinsic low symmetry of the wurtzite phase, whereas the latter results from the applied mechanical strain [71]. The origin of strain in the defected layers of thin films is usually the in-plane lattice mismatch $\Delta a/a$. Since all types of stacking faults are almost perfectly matched to the ideal structure (see Table 1), the piezoelectric polarization can be assumed to be negligible [25].

Within the Berry-phase [72] method, the spontaneous polarization is defined as the polarization difference between two different structures of the same solid: a polar non-centrosymmetric structure and its non-polar high-symmetry (reference) counterpart [73,74]. In our case, the polar non-centrosymmetric structure is wurtzite and the structure that serves as a reference is zincblende [75-78]:

$$P_{sp} = P_{WZ} - P_{ZB}, \quad (3)$$

where WZ and ZB denote the wurtzite and zincblende structures, respectively. Using the Berry-phase technique, the spontaneous polarization of each stacking fault is calculated and reported in Table 4 together with the available simulation data from the literature. The spontaneous polarization is found to be negative and non-zero only along the hexagonal *c*-axis, which corresponds to the ideal stacking direction.

It is important to observe the trend followed by the stacking faults: the spontaneous polarization decreases in magnitude with the number of cubic bonds, i.e. from type-I to type-II and, more significantly, to extrinsic stacking fault. Indeed, the latter is the closest structure to the non-polar zincblende phase (with three cubic bonds); consequently, its polarization is the smallest among the stacking faults studied here. Our values for the ideal structure (i.e. without any cubic bonds) agree well with the ones reported in the literature [77,78]. In addition, a self-consistent Poisson-Schrödinger calculation in Ref. [25] results in a value of -0.018 C/m² for the extrinsic stacking fault of GaN, which is very close to our value of -0.016 C/m².

**Table 4** Berry-phase results, namely spontaneous polarization $P_{sp}$ and interface polarization charge $\sigma$ for the ideal and stacking fault structures of wurtzite GaN and ZnO. Previous theoretical values [25,77,78,84] are given in parenthesizes. $\sigma$ has the same unit as polarization, i.e. C/m², but is expressed here in electrons per cm² (or simply cm⁻²). The conversion factor between the two units is 6.25×10¹⁴.

|  | Ideal | Type-I | Type-II | Extrinsic |
|---|---|---|---|---|
| **GaN** | | | | |
| $P_{sp}$ (C/m²) | -0.035 (-0.034[a]) | -0.026 | -0.021 | -0.016 (-0.018[b]) |
| $\sigma$ (10¹² cm⁻²) | 0 | 5.63 | 8.75 | 11.88 (11.08[c]) |
| **ZnO** | | | | |
| $P_{sp}$ (C/m²) | -0.044 (-0.057[d]) | -0.037 | -0.031 | -0.025 |
| $\sigma$ (10¹² cm⁻²) | 0 | 4.88 | 8.30 | 12.24 |

[a] Ref. [77], [b] Ref. [25], [c] Ref. [84], [d] Ref. [78]

As pointed out in Sec. III.2, stacking faults can be considered as quantum wells with zincblende/wurtzite interfaces. The spontaneous polarization differs for each side of the interface, thus giving rise to the accumulation of interfacial electric charge [79-81]. The main consequence of this charged layer is the built-in electric fields present at the stacking faults [82]. The interface charge $\sigma$ is equal to the difference between the spontaneous polarization of zincblende [with stacking faults (SF)] and wurtzite [without stacking faults (Ref)] structures [79-81]:

$$\sigma = P_{sp}^{SF} - P_{sp}^{Ref}. \quad (4)$$

This is often referred to as "the interface theorem" [83]. We calculate the interface charge using Eq. (4) and polarization data from Table 4 and present the results in the same table. We find that the interface charges for GaN stacking-faults are comparable to those for ZnO. The most charged interface is the one associated with the extrinsic stacking fault, which is consistent with the reported values of the spontaneous polarization. The agreement between our Berry-phase value and the band-structure result of Ref. [84] of the extrinsic-fault charge is very good, indicating the validity of both approaches in evaluating the polarization charge at interfaces. The polarization-induced charges and the ensuing electric fields considerably affect the performance of GaN- and ZnO-based applications. For example, the presence of polarization fields tends to reduce the optical efficiency, thereby limiting the performance of light-emitting devices [85,86].

**b) Piezoelectric coefficient**

Wurtzite GaN and ZnO are attractive piezoelectric materials for applications that require efficient electromechanical coupling [87,88]. As such, it is interesting to investigate the effect of stacking faults on the piezoelectric coefficients of both materials. Given that the hexagonal stacking direction coincides with the *c*-axis, we restrict ourselves to the axial piezoelectric constant $e_{33}$ that describes the polarization induced by a strain applied to the *c*-axis [78]. As emphasized in the Berry-phase theory and throughout this work, the piezoelectric coefficient is accessible via a finite difference of polarization between two states, corresponding to the strained and unstrained structures. Specifically, the piezoelectric coefficient is expressed as [89]:

$$e_{33} = \frac{\Delta P}{\varepsilon}, \qquad (5)$$

where $\Delta P$ refers to the difference of polarization before and after the application of the macroscopic strain $\varepsilon$. In order to determine the piezoelectric constants of the stacking faults, the corresponding wurtzite supercells are strained by -1% along the *c*-axis, then the polarizations are calculated as Berry phases. It is found that this value of the strain allows to reproduce the previously reported coefficients for perfect GaN and ZnO. The results of the piezoelectric coefficient for each type of stacking faults are displayed in Table 6. Comparing our results for the perfect crystals with previous calculations [90,91] shows a reasonable agreement.

**Table 6** Calculated piezoelectric coefficient $e_{33}$ for the ideal and stacking faults structures of wurtzite GaN and ZnO.

|  | Ideal | Type-I | Type-II | Extrinsic |
|---|---|---|---|---|
| **GaN** | | | | |
| $e_{33}$ (C/m²) | 0.67 (0.63-0.83)[a] | 0.62 | 0.56 | 0.51 |
| **ZnO** | | | | |
| $e_{33}$ (C/m²) | 1.10 (0.96-1.3)[b] | 1.06 | 1.02 | 0.98 |

[a] Ref. [90], [b] Ref. [91]

We note that the piezoelectric coefficient follows the same trend identified earlier for the spontaneous polarization: $e_{33}$ decreases in going from type I to type II and to extrinsic stacking fault. This implies that the piezoelectric performance decreases in the presence of these extended defects. It is therefore desirable to exploit materials with low concentration of stacking faults in the fabrication of electromechanical devices in order to improve their performance.

## IV. Summary

As a summary we present in Fig. 4 the formation energy of different stacking-faults versus the number of cubic layers for ZnO and GaN and observe a linear behavior of surprisingly high precision. Furthermore, the formation energy for stacking faults of type I coincides reasonably well with the energy difference between the zincblende and wurtzite polytypes as mentioned earlier. Let us repeat: the formation energy for different types of stacking faults in wurtzite crystals is proportional to the number of cubic layers. We have demonstrated it here for basal stacking faults in GaN and ZnO. However, the surprising precision of this simple rule of thumb leads us to conjecture that it is true in *any* wurtzite crystal (i.e. also for InN, AlN, BeO, or others) and probably also for other types of stacking faults.

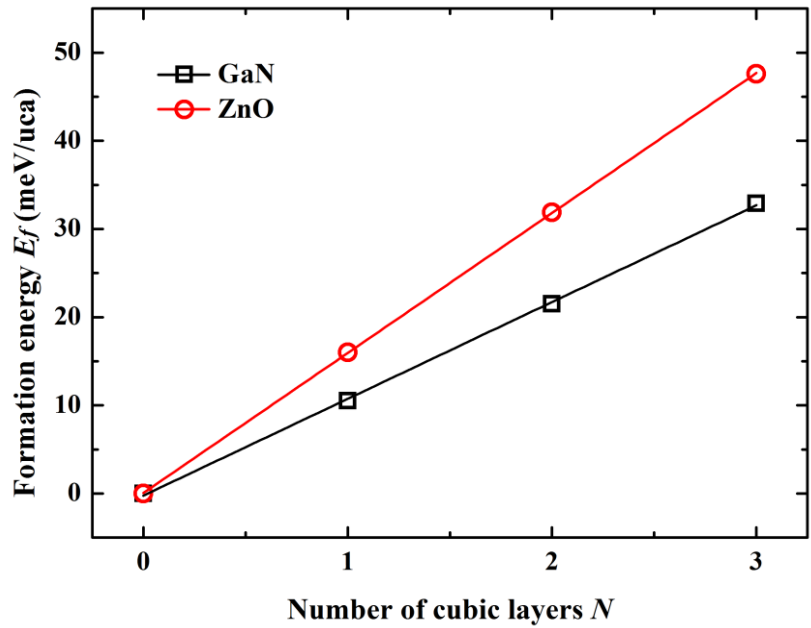

**Figure 4** Stacking-fault formation energies in meV per unit cell area (uca) versus the number of cubic layers in GaN and ZnO for each type of stacking faults. Curves are linear fits of the calculated data.

Similarly, the band gap energies and spontaneous polarizations for different concentrations of cubic layers in our chosen supercells are summarized in Figs. 5 and 6. The linear behavior is less evident than for the formation energies in Fig. 4 but still valid, especially for low concentrations of cubic layers. The deviations are larger for the supercell with 60 percent cubic layers (to simulate extrinsic stacking faults), both for the band gap and the polarization.

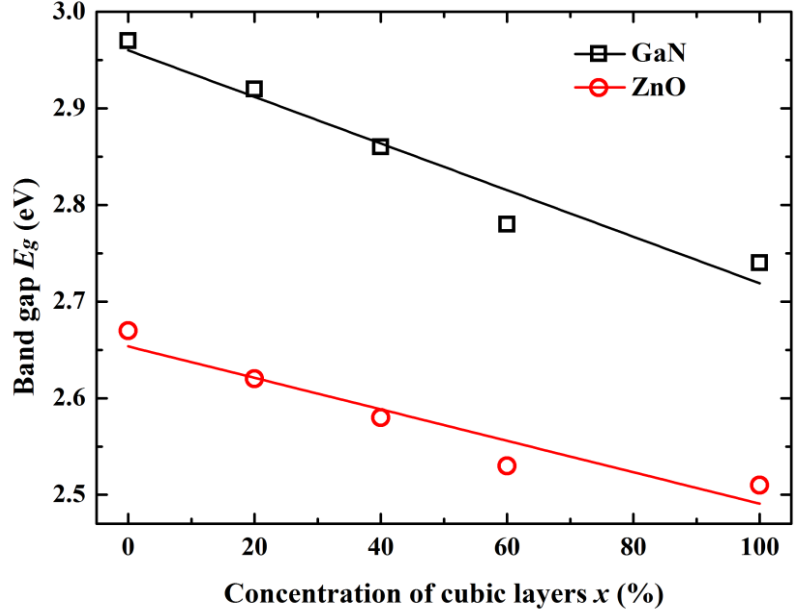

**Figure 5** Calculated mBJ band gap energies of GaN and ZnO for different concentrations of cubic layers $x$ in the chosen supercells. The concentrations $x = 0$ and 100 % correspond to the wurtzite and zincblende phases, respectively. Curves are linear fits of the calculated data.

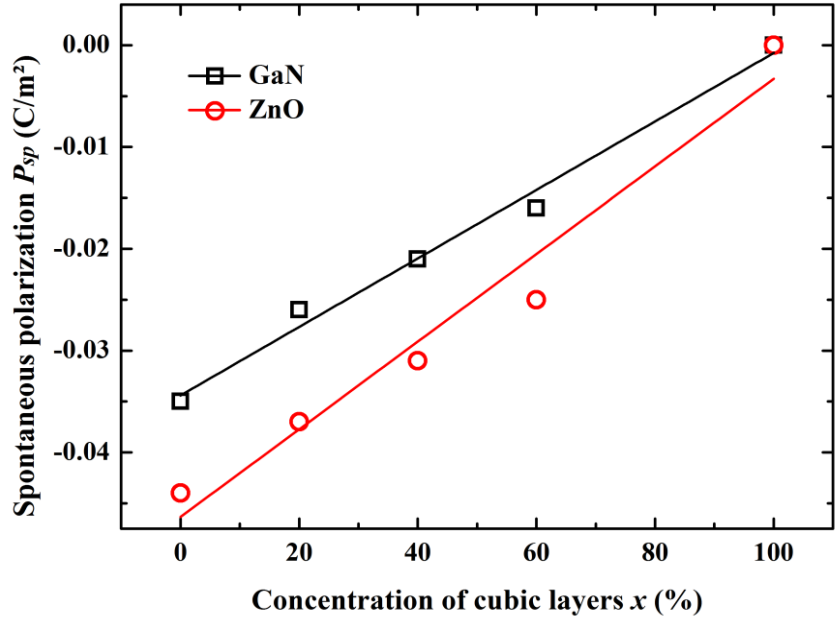

**Figure 6** Calculated spontaneous polarization of GaN and ZnO for different concentrations of cubic layers $x$ in the chosen supercells. The concentrations $x = 0$ and 100 % correspond to the wurtzite and zincblende phases, respectively. Curves are linear fits of the calculated data.

## V. Conclusion

In this paper, we have presented the results of a systematic study of basal-plane stacking faults in wurtzite GaN and ZnO, obtained from DFT calculations using periodic supercells. Specifically, we have investigated the crystal structure, electronic structure and electric polarization of the intrinsic and extrinsic stacking faults. Such investigation is of central importance since stacking faults are very common defects in real wurtzite materials. The overall message emerging from the present work is that stacking faults affect most of the properties of GaN and ZnO to different extents, as summarized below. From the structural relaxation, we find that stacking faults have a negligible influence on lattice parameters and atomic positions. The formation energies are quite low and increase in this order: type-I, type-II and extrinsic stacking fault. This ordering is consistent with the number of cubic layers at each type of defect and with the wurtzite/zincblende energy difference. The band-structure calculation reveals that the band gap of stacking faults is smaller compared to that of the perfect wurtzite phase, which means that these defects act as quantum wells whose height is determined by the valence and conduction band offsets. Both spontaneous polarization and piezoelectric coefficient decrease in magnitude when the stacking faults are introduced within the studied materials. In contrast, the polarization charge at the wurtzite/zincblende interface increases, fully consistent with the behavior of the spontaneous polarization. Taking all these results into consideration, we recommend the use of the values reported in this work in simulations involving basal stacking faults of wurtzite GaN and ZnO.